\documentstyle[aps,prb,twocolumn,epsfig,floats,amssymb,amsfonts]{revtex}

\def\s{\sum\limits}

\def\be{\begin{equation}}
\def\e{\end{equation}}
\def\beml{\begin{mathletters}}
\def\eml{\end{mathletters}}
\def\beq{\begin{eqnarray}}
\def\eq{\end{eqnarray}}
\def\ba{\begin{array}}
\def\a{\end{array}}
\def\d{\dagger}
\def\l{\left}
\def\r{\right}
\def\n{\nonumber}
\def\la{\langle}
\def\ra{\rangle}
\def\det{\,{\rm Det}\,}
\def\tr{{\rm Tr}\,}

\def\im{\,{\rm Im}\,}
\def\re{\,{\rm Re}\,}

\def\ep{\varepsilon}
\begin{document}
 
\draft

\date{May 2000}

\title{Can mesoscopic fluctuations
reverse the supercurrent through a disordered 
Josephson junction?}

\author{M. Titov, Ph. Jacquod, and C. W. J. Beenakker}

\address{
Instituut-Lorentz, Universiteit Leiden,
P.\,O.~Box 9506, 2300 RA Leiden,
The Netherlands
}
\twocolumn[
\widetext
\begin{@twocolumnfalse}

\maketitle

\begin{abstract}
We calculate the Josephson coupling energy 
$U_J(\phi)$ (related to the supercurrent
$I=(2e/\hbar) dU_J/d\phi$) for a disordered normal metal
between two superconductors with a phase difference $\phi$.
We demonstrate that the symmetry of the scattering matrix
of non-interacting quasiparticles in zero magnetic field
implies that $U_J(\phi)$ has a minimum at $\phi=0$.
A maximum (that would lead to a $\pi$-junction or negative
superfluid density) is excluded for any realization
of the disorder.
\end{abstract}

\end{@twocolumnfalse}
]
\narrowtext

The question in the title was posed ten years ago by Spivak
and Kivelson,\cite{SK} in their search for
mechanisms that would lead to a negative local superfluid density
in a dirty superconductor. A negative instead of a positive 
superfluid density means that the Josephson coupling energy 
$U_J(\phi)$ has a maximum instead of a minimum for zero
phase difference $\phi$ of the superconducting order parameter.
The supercurrent $I(\phi)=(2e/\hbar) dU_J/d\phi$ then has the opposite sign
as usual for small $\phi$. This has a variety of observable consequences,
including a ground state with a non-zero supercurrent,\cite{Bulaevskii}
Aharonov-Bohm oscillations with period $h/4 e$, and negative
magnetoresistance.\cite{KS} A Josephson junction with a negative superfluid density
is known as a $\pi$-junction,\cite{Bulaevskii} because the ground state for
large magnetic inductance is close to $\phi=\pi$ instead of 
as usual at $\phi=0$.

The known mechanism for the creation of a $\pi$-junction\cite{SK,Bulaevskii,KS,Glazman}
in equilibrium\cite{footnote1} in a non-magnetic material\cite{footnote2}
requires strong Coulomb repulsion to create a localized spin on a resonant impurity 
level. Spivak and Kivelson asked the question whether purely 
one-electron conductance fluctuations might be sufficient to produce a locally negative
superfluid density near the insulating state. A suggestive argument that
this might be possible comes from the relative magnitude
of the mesoscopic sample-to-sample fluctuations of the supercurrent in
a disordered superconductor--normal-metal--superconductor
(SNS) junction.\cite{Altshuler}
The ratio $\la \delta I^2 \ra^{1/2}/\la I\ra \simeq e^2/h G$ of the 
root-mean-squared fluctuations over the mean supercurrent is 
$\ll 1$ if the conductance $G$ of the normal metal is large
compared to the conductance quantum $e^2/h$. The ratio becomes 
of order unity on approaching the insulating state, suggesting
that the supercurrent might have a negative value in some samples.

Of course, the root-mean-square amplitude of the supercurrent fluctuations
does not
distinguish between a positive and negative sign, so that this argument 
is only suggestive. We were motivated to settle this issue
because of recent experiments on localization in quasi-one-dimensional 
superconductors.\cite{Bezryadin}
This has renewed the interest in the fundamental question whether mesoscopic 
fluctuations are sufficient or not to create a negative local superfluid
density. The answer, as we will show, is that they are not.

The two superconductors that form the SNS junction
have
order parameters $\Delta e^{i\phi/2}$ and $\Delta e^{-i\phi/2}$.
The contacts to the normal metal 
have $N$ propagating modes at the Fermi energy $E_F$, so that the elastic
scattering
by the normal metal at energy $E=E_F+\ep$ is characterized by a 
$2N\times 2N$
scattering matrix $S(\ep)$. The two properties of $S$ that we use
are that it is analytic in the upper half of the complex $\ep$ plane
and that it is a symmetric matrix, $S(\ep)=S(\ep)^T$,
when time-reversal symmetry is preserved. 

The starting point of our calculation is the relationship 
derived in Ref.\ \onlinecite{BB}
between 
the Josephson coupling energy  $U_J(\phi)$ in equilibrium 
at temperature $T$ and the scattering matrix,
\be
\label{basic}
U_J= -2 k_B T \s_{n=0}^\infty \ln \det 
\l[1-S_A(i\omega_n) S_N(i\omega_n) \r].
\e
The summation runs over the Matsubara frequencies 
$\omega_n= (2 n+1)\pi k_B T$.
The $4N\times 4N$ scattering matrix $S_N(\ep)$ describes
the elastic scattering by disorder in the normal metal
of non-interacting electron and hole quasiparticles
with excitation energy $\ep$,
\be
\label{SN}
S_N(\ep)=\l(
\ba{cc} S(\ep) & 0\\ 0 & S(-\ep)^* \a
\r).
\e 
The analytical continuation of $S_{N}$ from real 
to imaginary energy ($\varepsilon\rightarrow i\omega$) 
follows from $S(\varepsilon)\rightarrow S(i\omega)$ and
$S(-\varepsilon)^{\ast}\rightarrow S(i\omega)^{\ast}$.
Similarly, the matrix $S_A(\ep)$ describes the 
Andreev reflection from the superconductors,
\beml
\label{SA}
\beq
\label{SAa}
S_A(\ep)&=&\alpha(\ep)
\l( \ba{cc} 0 & e^{i \Lambda \phi/2} \\
e^{-i \Lambda \phi/2} & 0
\a \r),\\ 
\label{SAb}
\alpha(\ep)&=&\ep/\Delta-
i\sqrt{1-\ep^2/\Delta^2}.
\eq
\eml
Here $\Lambda$ is a $2N\times 2N$ diagonal matrix
with elements $\Lambda_{jj}=1$ for
$1\le j \le N$ and $\Lambda_{jj}=-1$ for $N+1\le j\le 2N$.

Eq.\ (\ref{basic}) differs from the usual 
representation of the Josephson energy as a sum over the
discrete spectrum $(\ep<\Delta)$ plus an integration over
the continuous spectrum $(\ep > \Delta)$. 
Each term in the sum (\ref{basic}) is combined from
contributions out of the discrete and the continuous spectrum.
We will now show that each of these combinations is minimal
for $\phi=0$, although the contributions from 
the discrete and continuous spectrum separately are not.

Let us abbreviate
\be
Z(\omega)=(\sqrt{1+\omega^2/\Delta^2}
-\omega/\Delta)e^{-i \Lambda \phi/2} S(i\omega).
\e
Using the identity $\ln \det =\tr \ln$ in Eq.\ (\ref{basic}) 
one can calculate the first and second derivative 
with respect to $\phi$ of the energy $U_J(\phi)$. 
The first derivative is given by
\be
\label{firsta}
{d U_J\over d\phi} = 2 k_B T \s_{n=0}^\infty \im \tr 
\l[ h_{11}(\omega_n) -h_{22}(\omega_n) \r],
\e
where the $N\times N$ matrices $h_{11}$ and $h_{22}$
are blocks of the matrix
\be
\label{firstb}
H=Z^* Z(1+Z^* Z)^{-1}=\l(
\ba{cc} h_{11}& h_{12}\\ h_{21} & h_{22} \a
\r).
\e
The first derivative
is equal to the supercurrent and vanishes at $\phi=0$ 
as dictated by time-reversal symmetry.

For the second derivative we obtain 
\beq
{d^2 U_J\over d\phi^2}& =& 4 k_B T \s_{n=0}^\infty \re \tr 
\l[ f_{12}(\omega_n) f_{21}^*(\omega_n) \r.\n \\
\label{seconda}
&& \l. +h_{12}(\omega_n) h_{21}(\omega_n) \r],\\
\label{secondb}
F&=&Z(1+Z^* Z)^{-1}=\l(
\ba{cc} f_{11}& f_{12}\\ f_{21} & f_{22} \a
\r).
\eq
At $\phi=0$ the symmetry of $S$ implies that 
$F=F^T$ and $H=H^\d$, hence  $f_{21}^*=f_{12}^\d$, $h_{21}=h_{12}^\d$.
Therefore, every term in the sum (\ref{seconda})
is positive. We conclude that the
Josephson energy
$U_J(\phi)$ has a minimum at $\phi=0$:
\be
\l.{dU_J\over d\phi}\r|_{\phi=0}=0, \qquad
\l.{d^2U_J\over d\phi^2}\r|_{\phi=0}>0.
\e

This concludes the proof that mesoscopic fluctuations can not invert the
stability of the SNS junction at zero phase, excluding a mechanism for
the creation of a $\pi$-junction proposed ten years ago.\cite{SK} 
The proof holds for non-interacting
quasiparticles in zero magnetic field at arbitrary temperature, for any
disorder potential and any dimensionality of the junction. As a final
remark, we conjecture (and have a proof for $N=1$) that Eq.\ (1) implies
$dU_{J}/d\phi>0$ in the entire interval $0<\phi<\pi$ in the presence of
time-reversal symmetry.

We thank Boris Spivak for urging us to solve this problem and for valuable
discussions. This work was supported by the Dutch Science Foundation NWO/FOM.

\end{document}